\documentclass{article}

\usepackage{authblk}

\usepackage[english]{babel}
\usepackage{tabularx}
\usepackage[letterpaper,top=2cm,bottom=2cm,left=3cm,right=3cm,marginparwidth=1.75cm]{geometry}
\usepackage[font=small,labelfont=bf]{caption}
\usepackage{amsmath}
\usepackage{graphicx}
\usepackage[colorlinks=true, allcolors=blue]{hyperref}
\usepackage{tikz}
\usepackage{amssymb}
\usepackage[math]{anttor}
\usepackage{caption}
\usepackage{subcaption}
\usepackage{xcolor}

\usetikzlibrary{shapes.geometric, arrows}
\tikzstyle{startstop} = [rectangle, rounded corners, minimum width=3cm, minimum height=1cm,text centered, draw=black, fill=red!30]
\tikzstyle{io} = [trapezium, trapezium left angle=70, trapezium right angle=110, minimum width=3cm, minimum height=1cm, text centered, draw=black, fill=blue!30]
\tikzstyle{process} = [rectangle, minimum width=3cm, minimum height=1cm, text centered, draw=black, fill=orange!30]
\tikzstyle{decision} = [diamond, minimum width=3cm, minimum height=1cm, text centered, draw=black, fill=green!30]
\tikzstyle{arrow} = [thick,->,>=stealth]

\title{Machine learning based data-driven discovery of nonlinear phase-field dynamics}

\author[2,3,5]{Elham Kiyani$^{1,2}$, Steven Silber$^{2,3}$, Mahdi Kooshkbaghi$^{4}$, Mikko Karttunen$^{2,3,5}$\\
  $^1$Department of Mathematics, The University of Western Ontario, 1151 Richmond Street, London, ON N6A~5B7 Canada\\
  $^2$The Centre for Advanced Materials and Biomaterials (CAMBR), The University of Western Ontario, 1151 Richmond Street, London, ON N6A~5B7 Canada\\
  $^3$Department of Physics and Astronomy, The University of Western Ontario, 1151 Richmond Street, London, ON N6A~3K7 Canada\\
$^4$Simons Center for Quantitative Biology, Cold Spring Harbor Laboratory, Cold Spring Harbor, NY, USA\\
$^5$Department of Chemistry, The University of Western Ontario, 1151 Richmond Street, London, ON N6A~5B7 Canada
}
\begin{document}

\maketitle
\begin{abstract}
One of the main questions regarding complex systems 
at large scales 
concerns the effective interactions and driving forces that emerge from the detailed microscopic properties.
Coarse-grained models 
aim to describe 
complex systems in terms of coarse-scale equations with a reduced number of degrees of freedom.
Recent developments in machine learning (ML) algorithms have significantly empowered the discovery process of the governing equations directly from data. 
However, it remains difficult to discover  partial differential equations 
(PDEs) with high-order derivatives.
In this paper, we present new data-driven architectures based on multi-layer perceptron (MLP), convolutional neural network (CNN), and a combination of CNN and long short-term memory (CNN-LSTM) structures for discovering the non-linear equations of motion for phase-field models with non-conserved and conserved order parameters.
The well-known Allen--Cahn, Cahn--Hilliard, and the phase-field crystal (PFC) models were used as the test cases.
Two conceptually different types of implementations were used:
(a) guided by  physical intuition (such as local dependence of the derivatives) and (b) in the absence of any physical assumptions (black-box model). We show that not only can we effectively learn the time derivatives of the field in both scenarios, but we can also use the data-driven PDEs to propagate the  field in time and achieve results in good agreement with the original PDEs.
\end{abstract}

\section{Introduction}

PDEs are widely used in modeling of complex physical, chemical and biological systems including fluid dynamics, chemical kinetics, population dynamics and phase transitions.
The study of PDEs in the context of ML, broadly speaking, falls into two categories: 1.) solving PDEs and 2.) predicting unknown PDEs from data \cite{han2018solving,712178,sirignano2018dgm, hure2019some,han2020algorithms,ranade2021discretizationnet,raissi2019physics}. 
In simulations of phase-field and reaction-diffusion models, commonly used numerical techniques are 
based on time and space discretization, such as finite difference and finite element methods. In recent years, a third approach based on ML has emerged with promising results for solving and even discovering unknown PDEs from data, see for example Ref.~\cite{karniadakis2021physics} and references therein.

The core idea for using ML algorithms 
to solve PDEs is representing the residuals of PDEs as a loss function of a neural network (NN) 
where the loss function is minimized;
a loss function measures how far the predicted values are from their true values.
This approach does not require discretization or meshing, which is beneficial when dealing with problems of high dimensions and/or complex geometries~\cite{han2018solving,712178, lu2021learning}. 
Since most deep learning frameworks are based on automatic differentiation, these methods are known as mesh-free approaches~\cite{lu2021deepxde}.

In the case of discovering unknown PDEs from data,
the key idea of ML-based approaches is to estimate the time derivative of the desired (dependent) quantity. These approaches can be broadly categorized as follows:
\begin{enumerate}
    \item An ensemble of macroscopic observations is available and there is knowledge about the physics of the governing coarse PDE(s). The typical knowledge is that the time evolution of the field of interest depends on the field and its derivatives (e.g. Navier-Stokes equations). One can design the ML algorithm to find that dependency. This relation can be formulated based any of the following methods: 
    (i) Linear dependence of the field evolution using a dictionary of spatial derivatives with unknown coefficients \cite{brunton2016discovering, schaeffer2017learning}; (ii) Nonlinear dependence with black-box inference \cite{raissi2019physics}; or (iii) Nonlinear dependence using a selective dictionary of spatial derivatives which were found by other data-driven approaches~\cite{lee2020coarse}; (iv) Nonlinear dependence where spatial derivatives are informed by the memory (history) of the system using a feedback loop, e.g.~Recurrent neural network (RNN) together with long short-term memory (LSTM) and gated recurrent unit (GRU)~\cite{vlachas2018data, gers2002applying, del2021learning}.
    \item An ensemble of microscopic observations is available and the macroscopic field of interest is known. For example, the microscopic solutions of the Boltzmann equation are available and one is looking for the time evolution of coarse fields such as density, velocity or temperature. Again, one can assume that the time evolution of the field depends on the spatial derivatives using physical intuition~\cite{bar2019learning,lee2020coarse}.
    \item An ensemble of microscopic observations is available but the macroscopic field is unknown. Therefore, the first step 
is to discover the coarse-grained field, which is generally formulated as a model reduction problem~\cite{theodoropoulos2006reduced}.
The second step is to find the PDE(s) for the coarse variable(s). 
For example, Thiem et al. determined an order parameter
for coupled oscillators
using diffusion maps and the corresponding governing PDE
using a Runge-Kutta network~\cite{thiem2020emergent}. 
\end{enumerate}
In this paper, we explore ML-based approaches which fall under the first category mentioned above. We assess two scenarios:
\begin{itemize}
    \item[(i)] There is an unknown relation between field evolution spatial derivatives.
    \item[(ii)] The spatial derivatives, their orders and combinations are unknown (there is no spatial derivative dictionary). 
\end{itemize}
Afterwards, we also solve the predicted PDEs in time and space.

For the first scenario, a flexible framework that can deal with large data sets and extract the unavailable PDE(s) from coarse-scale variables implicitly is developed. Two different approaches are presented for learning coarse-scale PDEs: 1) an MLP architecture and 2) a CNN-LSTM. Since LSTM only passes time information to its layers and 
misses the spatial features of previous time steps, CNN can be used to learn and detect the spatial features of the inputs~\cite{kim2014convolutional,8632898}.
For the second scenario, a convolution operator is used to implicitly learn the dependence of the time derivative of the field on the spatial derivative(s) of unknown orders. The learned PDE is then marched in time with a time-integrator.

We demonstrate the capability of the above algorithms to learn PDEs 
using data
obtained from phase-field simulations 
of the well-known
Allen--Cahn~\cite{allen1972ground}, Cahn--Hilliard~\cite{cahn1958free}, and the phase-field crystal (PFC)~\cite{Elder2002,Elder2004-jc} models 
using the open source software \textit{SymPhas}~\cite{silber2021symphas}.

The rest of this article is organized as follows.  A brief summary of the phase-field approach
and data preparation is presented in Section~\ref{Phase-field model}. In Section~\ref{Data-Driven PDEs with Spatial Derivatives Dictionary}, an overview of MLP and CNN-LSTM networks as two data-driven approaches which learn PDEs with a spatial derivatives dictionary is presented.
Finally, in Section~\ref{CNN}, we introduce a CNN network that learns PDEs without any assumption regarding spatial derivatives. Then, the solutions of the original and data-driven PDEs are compared. 
The \texttt{tensorflow2} framework \cite{tensorflow2015-whitepaper} was used to implement and train our networks.

\section{Phase-Field Modeling}\label{Phase-field model}

\subsection{Phase-Field Modeling in a Nutshell}

Phase-field modeling provides a theoretical and computational approach for simulating
non-equilibrium processes
in materials, typically with the objective of studying the dynamics and structural changes. For an excellent overview of phase-field modelling, see the book by Provatas and Elder~\cite{Provatas2010-hc}. In its essence, phase-field modeling is a coarse-grained approach that uses continuum fields to describe slow variables such as concentrations instead of atoms. The continuum fields are given by order parameters which may be conserved or non-conserved, and the dynamics of the order parameter(s) is (are) described by  a Langevin equation, or equations when several order parameters are present.
In this work, we consider only systems described by a single order parameter $U \equiv U(\vec{x}, t)$. In all of the descriptions below, we use the conventional dimensionless units~\cite{Provatas2010-hc}.

The Langevin equations, or equations of motion, for the non-conserved and conserved order parameters are (see Ref.~\cite{Provatas2010-hc} for a more detailed discussion)
given as 
\begin{align}
    \dfrac{\partial U}{\partial t} &= -\Gamma \frac{\delta F}{\delta U} & \text{(non-conserved)}& \label{eq:pf-nonconserved}\\
    \dfrac{\partial U}{\partial t} &= \Gamma \nabla^2\frac{\delta F}{\delta U} & \text{(conserved),}& \label{eq:pf-conserved}
\end{align}
where $F$ is a free energy  functional and $\delta / \delta U$ is a functional derivative. We have neglected thermal noise from the above equations. The parameter $\Gamma$ is a generalized mobility and is assumed to be constant. It is also noteworthy that when no free energy functional is available, the equations of motion are often postulated. This is the case with reaction-diffusion systems, for example, the well-known Turing~\cite{Turing1952,Leppanen2002-md} and Gray--Scott models~\cite{Gray_1985,Leppanen2002-md}.
Phase-field models have been widely applied to various types of systems and phenomena, including 
dendritic and directional solidification \cite{Grossmann_1993, Boettinger2002, Nestler_2005},
crystal growth \cite{Elder2002, Heinonen2016, Alster2020, Provatas2007} and
magnetism~\cite{Faghihi2019-uv}
as well as for phenomena such as fracture propagation \cite{Aranson_2000, Spatschek_2011} to mention some examples.

\subsection{Phase-Field Models used in the Current Work}

We employed three different well-studied single order parameter phase-field models: 1) The Allen--Cahn model \cite{allen1972ground} for the case of a non-conserved order parameter, 2) the Cahn--Hilliard model \cite{cahn1958free} for the conserved order parameter, and 3) the phase-field crystal (PFC) model \cite{Elder2002} that has a conserved order parameter and generates a modulated field that describes atomistic length scales and diffusive times. In addition to being well-studied, these models were chosen because they contain differing orders of spatial derivatives: the Allen--Cahn model is described using a 2nd order derivative, Cahn--Hilliard using 4th, and PFC using 6th. Moreover, they each exhibit various spatial patterns that evolve according to different time scales.

\subsubsection{The Allen--Cahn Model}

The Allen--Cahn model, a dynamical model for solidification originally developed by Allen and Cahn, 
has a single non-conserved order parameter 
corresponding to Equation~\eqref{eq:pf-nonconserved}, and is defined using a free energy density with a double well potential. The equation of motion has the form
\begin{equation}
\dfrac{\partial U}{\partial t} =  -M\left(\nabla^{2} U + a_{2} U + a_{4 }U^3\right)\,, \label{eq:modela}
\end{equation}
where $M = \Gamma$ is related to chemical mobility.
It represents a physical system that evolves purely due to a chemical potential. It is also called Model A according to the Hohenberg and Halperin classification of phase-field models \cite{hohenberg1977theory}. 

\subsubsection{The Cahn--Hilliard Model}

The Cahn--Hilliard model, formulated by Cahn and Hilliard,  represents spinodal decomposition.
It is a conserved order parameter model corresponding to Equation~\eqref{eq:pf-conserved}. Like the Allen--Cahn model, it is also defined using a free energy density corresponding to a double well potential.  The Cahn--Hilliard model and has the form
\begin{equation}
\dfrac{\partial U}{\partial t} =   D \nabla^2 \left(\nabla^{2} U + a_{2} U + a_{4 }U^3\right)\,, \label{eq:modelb}
\end{equation}
where $D = \Gamma$ represents the diffusion constant. 
It is also known as Model B in the Hohenberg and Halperin classification \cite{hohenberg1977theory}.

\subsubsection{The Phase-Field Crystal Model}

A free energy density %
to describe a crystal lattice at an atomistic scale incorporating elastic effects into a phase-field model was originally developed by Elder et al.~\cite{Elder2002,Elder2004-jc}. The free energy is minimized by a hexagonal periodic lattice, and is given by
\begin{equation}
    f(U) =\nabla ^{2}\left(\frac{U^3}{3} + \frac{U^4}{4} + U\left((q_{0} + \nabla ^{2})^{2}-\varepsilon\right)\frac{U}{2}\right),
\end{equation}
where $q_0$ and $\varepsilon$ are constants.
The equation of motion 
with a conserved order parameter can then be written as
\begin{equation}
    \dfrac{\partial U}{\partial t}=\nabla ^{2}\left(U^2 + U^3 + \left((q_{0} + \nabla ^{2})^{2}-\varepsilon\right)U\right)\,. \label{eq:pfc}
\end{equation}
The order parameter, $U$, represents the mass density.
The PFC model can used to describe elastic and plastic deformations in isotropic materials, i.e., crystal structures. The lattice structure
can assume any orientation 
(based on the initial conditions) and interactions of different grains (individual crystal structures) with each other can lead to defects and dislocations.

\subsection{Simulation of Phase-Field Models}

\begin{figure}
    \centering
    \includegraphics[width=\textwidth]{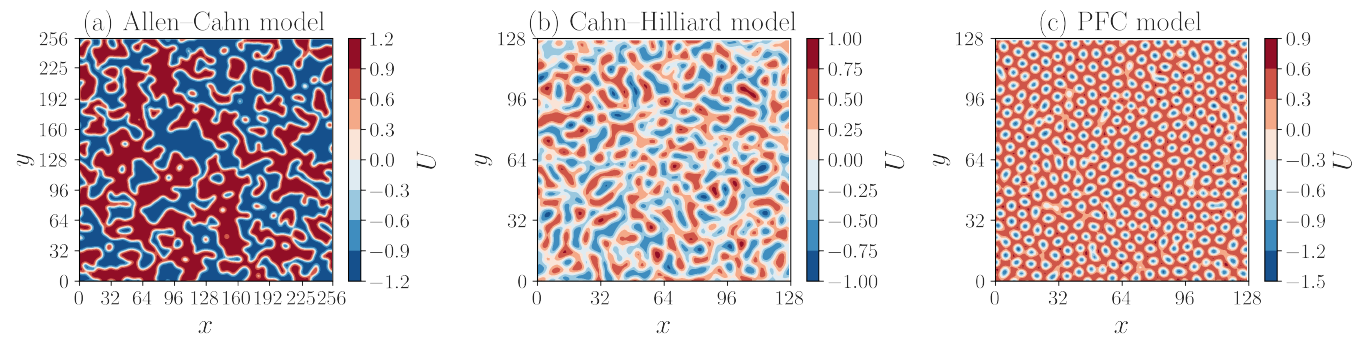}
    \caption{Snapshots from the three 
    phase-field models at various times. Field solutions for the Allen--Cahn model (Equation~\eqref{eq:modela}) is shown on the left at $t=20$, for the Cahn--Hilliard (Equation~\eqref{eq:modelb}) in the center at $t=20$, and for the PFC (Equation~\eqref{eq:pfc}) on the right at $t=100$. The parameters of the numerical simulations
are presented in Table~\ref{tab:input_data}. The vertical and horizontal axes display $x=n_x$ and $y=n_y$ respectively, and $U$ represents the phase-field for the corresponding
    model. These are dimensionless units.
    }
    \label{fig:model-examples}
\end{figure}

The open-source \textit{SymPhas} \cite{silber2021symphas} software package was used to numerically simulate the above three systems. 
\textit{SymPhas} 
allows the user to define phase-field models directly from their PDE formulations, and control simulation parameters from a single configuration file. All simulations were done using dimensionless units~\cite{Provatas2010-hc}.
In terms of the numerical solution, \textit{SymPhas} has the ability to simulate models using either explicit finite difference methods or the semi-implicit Fourier spectral method. The latter 
was chosen.
By virtue of its excellent error properties, the Fourier semi-implicit spectral method typically allows for larger time stepping 
than a finite difference solver~\cite{Chen1998}. 
\begin{table}
\centering
\begin{tabular}{l|l|l|l|l}
\hline
Phase-field model      & $n_x \times n_y$  & $\Delta t$ & $t$ & Equation parameters\\ 
\hline
Allen--Cahn, Eq~\eqref{eq:modela}    & $256 \times 256$  & 0.1 & 20 & $a_2 = a_4 = 1$ \\ 
Cahn--Hilliard, Eq~\eqref{eq:modelb} & $128 \times 128$  & 0.01 & 20 & $a_2 = a_4 = 1$ \\ 
PFC, Eq~\eqref{eq:pfc}           & $128 \times 128$  & 0.05 & 100 & $q_0 = 1$ and $\varepsilon = 0.1$ \\ \hline
\end{tabular}
\caption{
The simulations were done on a uniformly discretized square grid of side length $n_x \times n_y$ and spatial width $h = 1$. The simulations use a time step of $\Delta t$ and continue to time $t$. All models use periodic boundary conditions, and initial conditions were generated using a uniform random distribution. 
}
\label{tab:input_data}
\end{table}
For each of the models, five independent simulations
with 100 frames of the field $U$ were saved at equally spaced intervals.
Parameters for the numerical simulations are
summarized in Table~\ref{tab:input_data}. 
Each model uses a different time step or final time, and these were qualitatively chosen based on how long each model requires to reach a characteristic pattern. 
To illustrate the different dynamics of each model, snapshots at the end of each simulation are provided in Figure~\ref{fig:model-examples}.

\section{Data-Driven PDEs with Spatial Derivatives Dictionary}\label{Data-Driven PDEs with Spatial Derivatives Dictionary}

As already discussed above, we consider two distinct types of methods for discovering PDEs from data, assuming the network is informed by spatial derivatives either explicitly or implicitly.
In the first method, an MLP network is used to learn a function $F_{\mathrm{MLP}}$ that can be formulated as
\begin{equation}\label{eq:equation1}
U_{t}(t,x,y) = F_{\mathrm{MLP}}\left(U(t,x,y),U_{x}(t,x,y),U_{xx}(t,x,y), U_{y}(t,x,y), U_{yy}(t,x,y), ...\right), 
\end{equation}
where $U_{t}(t,x,y)$ is the time derivative and $U_{x}(t,x,y), U_{xx}(t,x,y), U_{y}(t,x,y)$ and $U_{yy}(t,x,y)$ are the first and second spatial derivatives with respect to $x$ and $y$, respectively.

In the second method, extending LSTM to a convolutional structure (CNN-LSTM) is used to learn an equation from local variables without giving spatial derivatives explicitly.
Mathematically, the network learns the time derivative $U_{t}(t,x,y)$ as a function of local macroscopic variables on a small square around each grid point,
\begin{equation}\label{eq:equation2}
U_{t}(t,x_{i},y_{j}) = F_{\mathrm{CNN-LSTM}}\left(U(t,x_{i-1},y), U(t,x_{i},y_{j}), U(t,x_{i+1},y_{j}), U(t,x_{i},y_{j-1}),  U(t,x_{i},y_{j+1})\right),
\end{equation}
where $U(t,x_{i-1},y_{j})$, $U(t,x_{i},y_{j})$, $U(t,x_{i+1},y_{j})$, $U(t,x_{i},y_{j-1})$, and $U(t,x_{i},y_{j+1})$ are the field values at the local positions $x_{i-1}$, $x_{i}$, $x_{i+1}$, $y_{j-1}$, and $y_{i+1}$, respectively.

A schematic diagram of our framework for discovering PDEs with spatial derivatives dictionary is shown in Figure~\ref{fig:CNN-LSTM-Schematic}. 
Specifically, it shows how spatial derivatives $(U, U_{x}, U_{y}, U_{xx}, U_{yy},...)$ and local macroscopic variables $(U_{i-1,j}, U_{i,j}, U_{i+1,j}, U_{i,j-1}, U_{i,j+1})$ are fed through the MLP (Figure~\ref{fig:CNN-LSTM-Schematic}a) and CNN-LSTM (Figure~\ref{fig:CNN-LSTM-Schematic}b), respectively, to learn the time derivative $U_{t}(t,x,y)$.

\subsection{Multi-layer Perceptron Network Architecture and Performance}\label{sec:LSTM}
An MLP is an example of a typical feedforward artificial neural network, consisting of a series of layers. Each layer calculates the weighted sum of its inputs and then applies an activation function to get a signal that is transferred to the next neuron~\cite{barlow1995feed}. 

In our MLP network, the number of layers, neurons, and the type of activation functions for each phase field models were found by trial and error. We approximate spatial derivatives of the coarse variable $U$ by finite differences, and along with $U$ itself, feed this to the MLP network to learn the function $F_{\mathrm{MLP}}$ in Equation~\eqref{eq:equation1}. 

\begin{figure}
    \centering
    \includegraphics[width=1\textwidth]{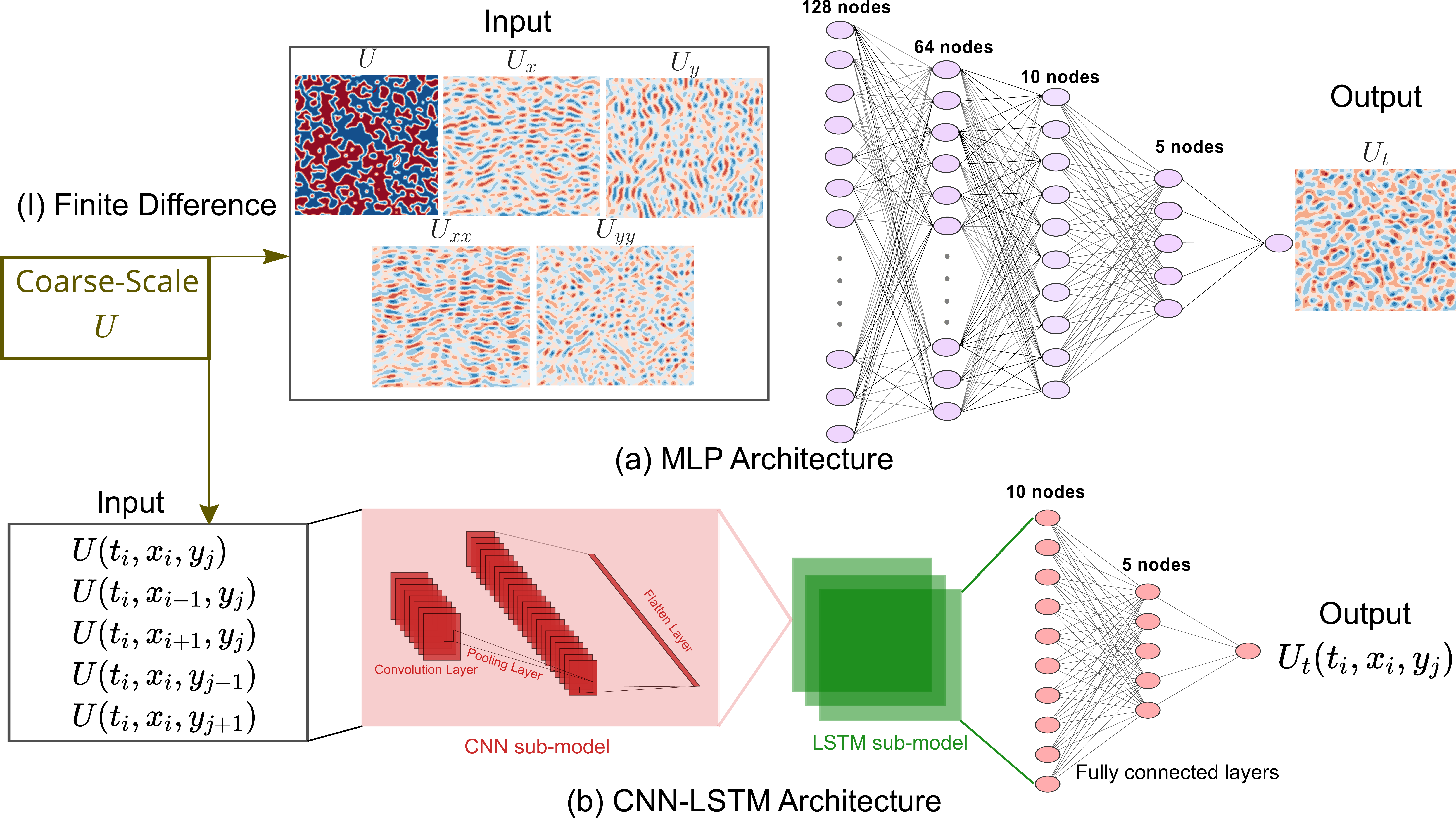}
    \caption{
    Schematic of the general steps in discovery of PDEs with spatial derivatives dictionary.
Learning of PDEs from spatial derivatives and local values of coarse variables using two different approaches, (a) MLP and (b) CNN-LSTM.
Coarse-scale variables are collected as snapshots from phase-field simulations. (I) Finite difference methods are used to approximate the spatial derivatives which are fed into (a) the MLP network according to Equation~\eqref{eq:equation1}. The network connecting the input layer consists of a list of input features (the field $U$ and its spatial derivatives) to the output layer of a single neuron (time derivative $U_{t}$).
    (II) The values of the macroscopic field $U$ evaluated 
    around each grid point are fed through (b) the CNN-LSTM network to learn the PDEs of the form given in Equation~\eqref{eq:equation2}. CNN-LSTM network connecting the input layer consists of a list of input features (local variables $U(t,x_{i-1},y_{i}), U(t,x_{i},y_{i}), U(t,x_{i+1},y_{i}), U(t,x_{i},y_{i-1}), U(t,x_{i},y_{i+1})$) to the output layer of a single neuron (time derivative).
}
    \label{fig:CNN-LSTM-Schematic}
\end{figure}

The MLP architecture for learning the Allen--Cahn model (Equation~\eqref{eq:modela}) is shown in Figure~\ref{fig:CNN-LSTM-Schematic}a) as an example. The input layer consisting of five neurons $U(t,x,y)$, $U_{x}(t,x,y)$, $U_{xx}(t,x,y)$, $U_{y}(t,x,y)$ and $U_{yy}(t,x,y)$, and is connected to the hidden layers which have 128, 64, 16, and 8 neurons each. In the output layer, we use a dense layer with a single neuron to predict $U_t$. The network is trained for $2,000$ epochs using the ADAM optimizer \cite{diederik2014adam} and rectified linear unit (ReLU) activation function \cite{brownlee2019gentle}. For the Cahn--Hilliard (Equation~\eqref{eq:modelb}) and PFC (Equation~\eqref{eq:pfc}) models, we used spatial derivatives up to fourth and sixth order for the input layers, respectively.

\begin{figure}
\begin{subfigure}{\textwidth}
  \centering
  \includegraphics[width=\linewidth]{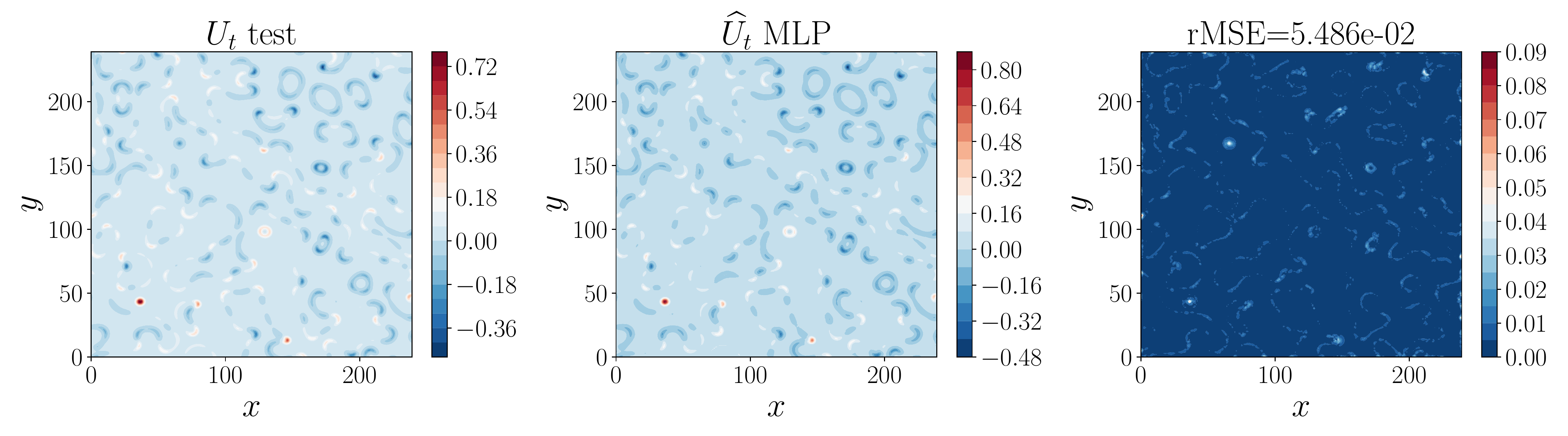}  
  \caption{Allen--Cahn model}
  \label{fig:modelA-MLP}
\end{subfigure}
\newline
\begin{subfigure}{\textwidth}
  \centering
  \includegraphics[width=\linewidth]{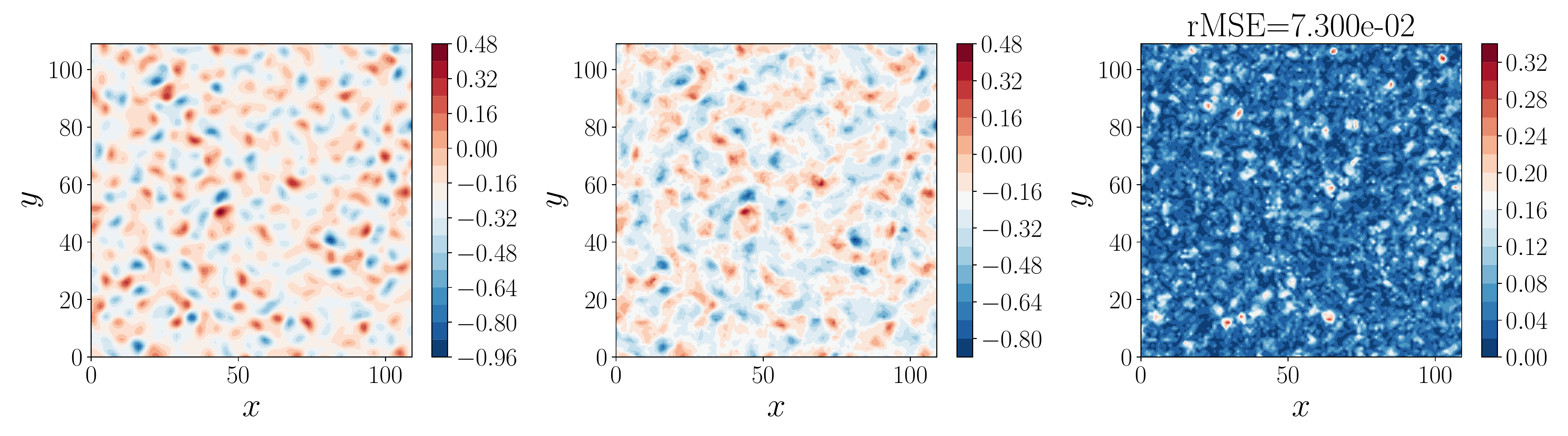}  
  \caption{Cahn--Hilliard model}
  \label{fig:moselB-MLP}
\end{subfigure}
\newline
\begin{subfigure}{\textwidth}
  \centering
  \includegraphics[width=\linewidth]{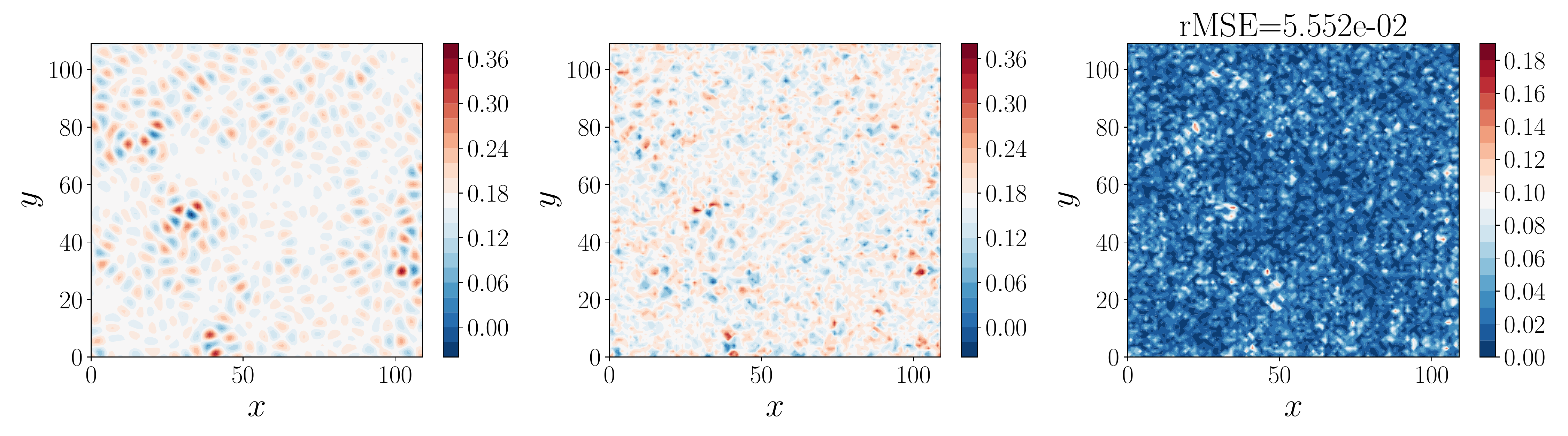}  
  \caption{PFC model}
  \label{fig:PFC-MLP}
\end{subfigure}
\caption{Performance of the MLP network on the prediction of the phase-fields given in Equations~\eqref{eq:modela}, \eqref{eq:modelb}, and \eqref{eq:pfc}. In each contour, $x=n_x$, $y=n_y$. The first contour shows $U_t$, the time derivative computed from the numerical solution generated by \textit{SymPhas}~\cite{silber2021symphas}, and the center contour shows $\widehat{U_t}$, the learned time derivative.  The right panel shows the difference between $U_t$ and $\widehat{U_t}$, as well as the corresponding rMSE value for each phase-field model. Our data is randomly split 60:20:20 into training, validation, and test sets. The network is trained for $2,000$ epochs using the ADAM optimizer and ReLU activation function.
}
\label{fig:LSTM1}
\end{figure}

The performance of the MLP network on learning 
the models is shown in Figure~\ref{fig:LSTM1}. The root mean squared error (rMSE) is the square root of the mean squared error (MSE) calculated as 
\begin{equation}\label{eq:MSE}
     \mathrm{MSE} = \frac{1}{n_x\times n_y} \sum_{i=1}^{n_x\times n_y}(U^i_{t} - \widehat{U_t^i})^2.
\end{equation}
As shown in Figure~\ref{fig:LSTM1}, the rMSE values are small ($\sim \! 10^{-2}$) indicating that the target time derivatives ($\widehat{U_{t}}$) learned by the proposed MLP network are close to the true ones for all three models.

\subsection{Convolution
and Long Short-Term Memory (CNN-LSTM)  Network Architecture and Performance}\label{LSTM and CNN-LSTM}

One of the main challenges in approximating coarse-scale PDEs is the estimation of  spatial derivatives.
While in previous studies PDEs have been successfully identified by learning time derivatives as a function of the estimated spatial derivatives, approximating derivatives remains challenging~\cite{LZIEGLER198777,press1991numerical}.
Generally, the choice of the time step is one of the most important considerations in numerical differentiation. While large step sizes can increase simulation speed, too large steps can create instabilities. On the other hand, if the steps are too small, numerical errors can dominate and the derivatives are of no use. Accordingly, the question that arises in discovering PDEs is the accuracy of numerical differentiation that has been used for training.

Unlike an MLP, CNN-LSTM is capable of automatically learning time derivatives from coarse-scale values. Using a combination of convolutional layers with other network structures for data-driven differential equations is an active field of research (see, for example, Refs.~\cite{2021, 2020}). CNNs are widely used for image classification and there have been several breakthroughs in image recognition with performance close to that of humans~\cite{szegedy2015going}.
The CNN architecture can progressively extract higher level representations (color, shape, topology, etc.) of an input feature (image) and learn the dependency of the output (mostly a single class label) to those representations. The convolution operation sweeps a filter across the entire input field and extracts the global features and local (pixel-to-pixel) variations. 
The convolutional layer in the networks can be considered as an efficient implementation of the convolution operator, hence, this layer represents approximations of (potentially of high order) derivatives of a scalar field. The convolution-differentiation connection and derivatives-order of filters relationships have been discussed in detail by Cai and Dong~\cite{cai2012image,dong2017image}. For particular applications where the desired outputs include localization (a class label is assigned to each pixel), a specific CNN architecture called ``U-net'' has been proposed~\cite{ronneberger2015u}. Since in most engineering and physics applications the time evolution of the scalar field depends on the local spatial derivatives, the U-net architecture is a reasonable candidate for such a learning task. The U-net-inspired network has also been successfully used in subgrid flame surface density estimation for premixed turbulent combustion modeling~\cite{lapeyre2019training}.
\begin{figure}
\begin{subfigure}{\textwidth}
  \centering
  \includegraphics[width=\linewidth]{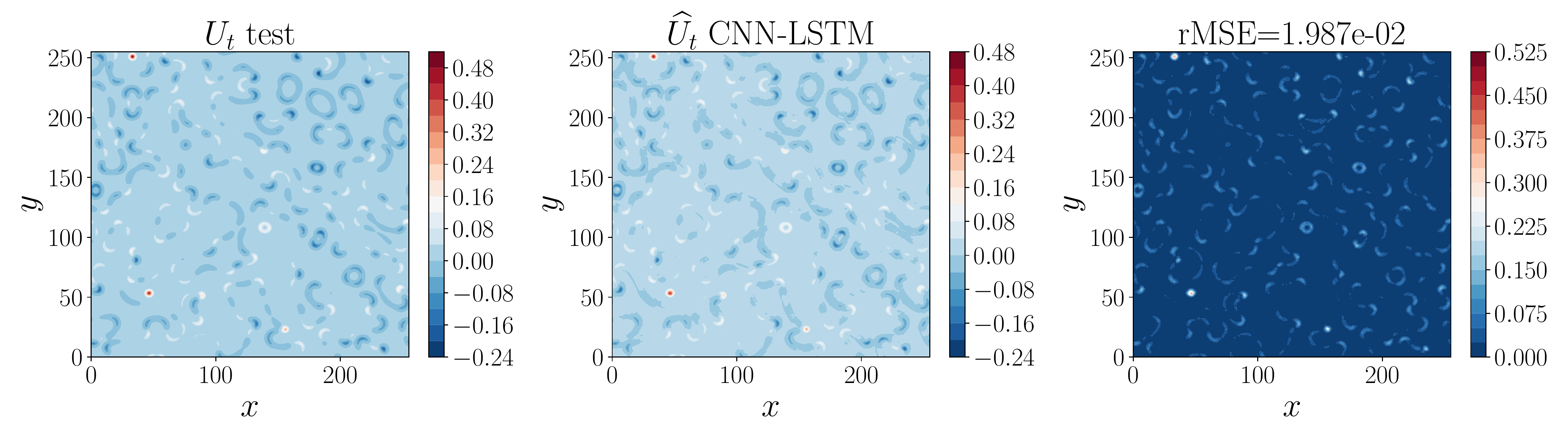}  
  \caption{Allen--Cahn model}
  \label{fig:CNN-LSTM-Allen--Cahn}
\end{subfigure}
\newline
\begin{subfigure}{\textwidth}
  \centering
  \includegraphics[width=\linewidth]{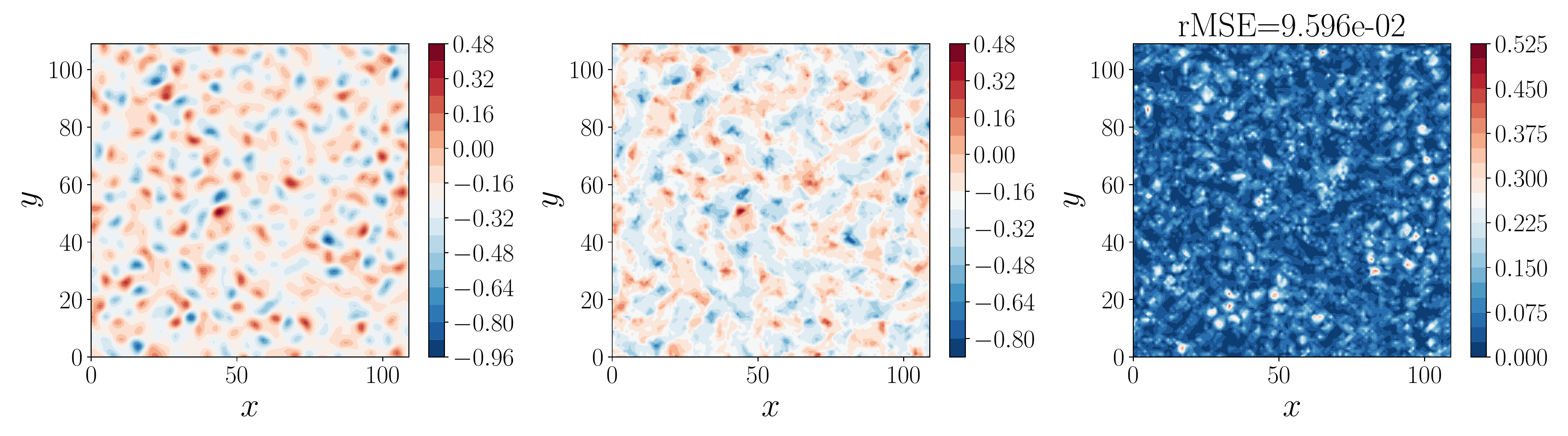}  
  \caption{ Cahn--Hilliard model}
  \label{fig:CNN-LSTM -Cahn--Hilliard}
\end{subfigure}
\newline
\begin{subfigure}{\textwidth}
  \centering
  \includegraphics[width=\linewidth]{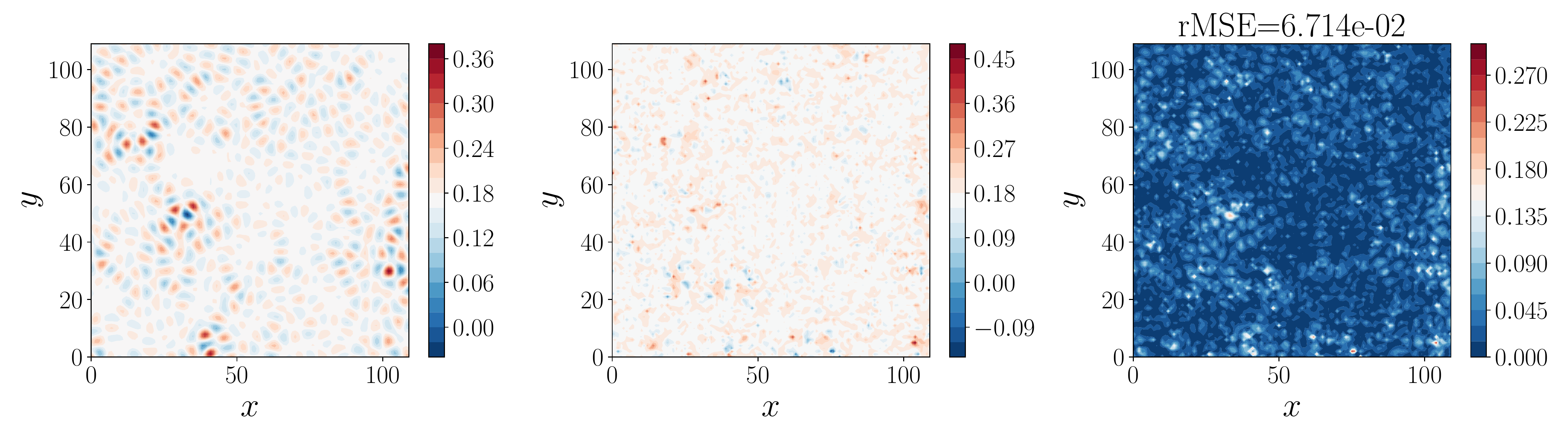}  
  \caption{PFC model}
  \label{fig:CNN-LSTM-pfc}
\end{subfigure}
\caption{CNN-LSTM predictions for  (a) the Allen--Cahn Equation~\eqref{eq:modela}, (b) the Cahn--Hilliard Equation~\eqref{eq:modelb}, and (c) the PFC Equation~\eqref{eq:pfc}. Actual and learned time derivatives $U_{t}$ and $\widehat{U_t}$ are shown in the left two panels. The difference between the predicted and the actual time derivatives as well as rMSE are presented in the right panel. Our data is randomly split 60:20:20 into training, validation and test sets, and the network is trained for 2,000 epochs using the ADAM optimizer and ReLU activation function.}
\label{fig:CNN-LSTM}
\end{figure}

A schematic diagram of the proposed CNN-LSTM architecture is shown in Figure~\ref{fig:CNN-LSTM-Schematic}b). The architecture consists of two sub-networks: (i) A CNN sub-network, including one-dimensional convolution and maxpooling layers for feature extraction from input data and, (ii) a LSTM sub-network including sequential layers followed by one LSTM layer and two dense layers with ReLU activation. We feed the CNN-LSTM network with the local coarse-scale variables, 
i.e., $(U(t_{i},x_{i-1},y_{i})$, $U(t_{i},x_{i},y_{i})$, $U(t_{i},x_{i+1},y_{i})$, $U(t_{i},x_{i},y_{i-1}), U(t_{i},x_{i},y_{i+1}))$ tuple. These coarse-scale variables at each grid point are fed into the CNN sub-network and the output of the convolutional layer passes through the LSTM layer followed by a dense layer to provide the output. The output of network is the single neuron approximating $U_t(t,x,y)$ at each grid point.

The LSTM network consists of a cell state which is the core concept of LSTM networks and memory blocks. Each block is composed of gates that can make decisions about which information passes through the cell state and which information can be removed. There are three kinds of gates: 1) input, 2) output, and 3) forget gate. Each memory block in an LSTM architecture has an input and an output gate which control information coming into the memory cell and information going out to the rest of the network, respectively. In addition, an LSTM architecture has a forget gate which contains an activation function and allows the LSTM to keep or forget information. Information from the previous hidden state and information from the current input is passed through the activation function. The output of each gate is a value between $0$ (block the information) and $1$ (pass the information)~\cite{6795963,Graves2012}.

The performance of the trained CNN-LSTM network is shown in Figure~\ref{fig:CNN-LSTM}. The network consists of a convolutional layer with 64 filters before pooling layer, kernel size 3 followed by an LSTM layer with 80 neurons. There are two dense layers (fully connected) with 10 and 5 neurons each.
In the left two panels, contours of ${U_t}$ and $\widehat{U_t}$ for the test sets and the corresponding predictions  by CNN-LSTM are compared. The difference between the original and the predictions are shown in the right panels, where rMSE is also reported for each phase-field model. The prediction errors from CNN-LSTM remain unchanged and an rMSE $\sim \! 10^{-2}$ is obtained for all three models.

\subsection{Hyper-parameter study}\label{comparison}

\begin{figure}
     \centering
     \includegraphics[width=\textwidth]{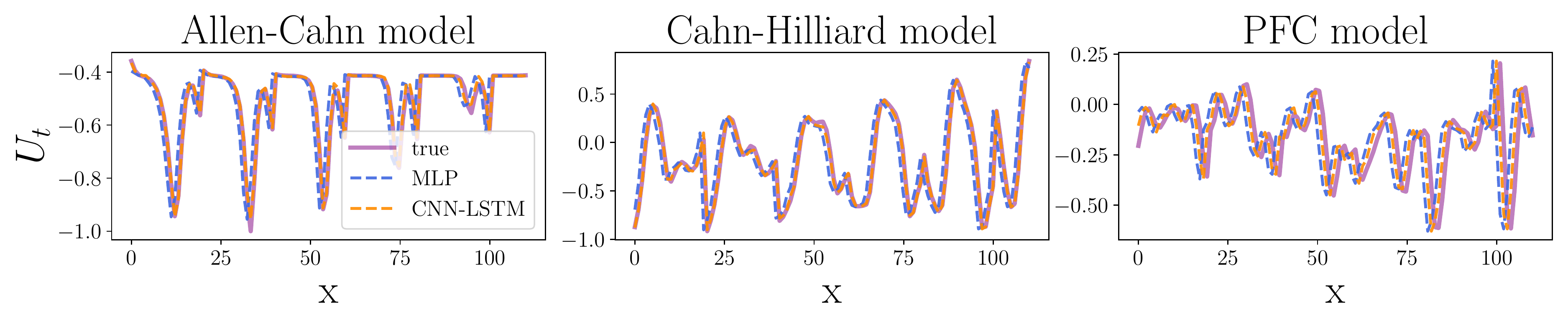}
     \caption{
              Comparisons between MLP and CNN-LSTM performance using the phase-field Equations~\eqref{eq:modela}, \eqref{eq:modelb}, and \eqref{eq:pfc}. In each plot, the vertical axis indicates $x=nx$ and the horizontal axis represents the time derivative $U_t$ and $\widehat{U_t}$ predicted by MLP and CNN-LSTM for each phase-field model. Two MLP and CNN-LSTM networks are trained and tested on the same data sets.
 }
      \label{fig:Comp-LSTM-CNN-LSTM}
 \end{figure}

A comparison of regression results over the selected prediction period obtained by 
MLP and CNN-LSTM is shown in Figure~\ref{fig:Comp-LSTM-CNN-LSTM}. One can clearly see the ability of both MLP and CNN-LSTM to accurately reproduce the original data and make predictions of the phase-field models.

We used the coefficient of determination, $R^2$ to compare the performance of the networks,
\begin{equation}\label{eq:R2}
R^2 = 1- \frac{\sum_{i=1}^{n_x\times n_y}(U^i_{t} - \widehat{U_t^i})^2}{\sum_{i=1}^{n_x\times n_y}(U^i_{t} - \overline{U_t})^2},
\end{equation}
where $\overline{U_t}$ is the mean value of the time derivative for a single snapshot.
Root mean squares and $R^2$ scores can be affected by different hyper-parameters such as learning rate, number of training epochs, and network depth and width. Here, we study the effect of adding/removing MLP and convolutional layers while all the other parameters are fixed.

\begin{figure}
\begin{subfigure}{\textwidth}
  \centering
  \includegraphics[width=\linewidth]{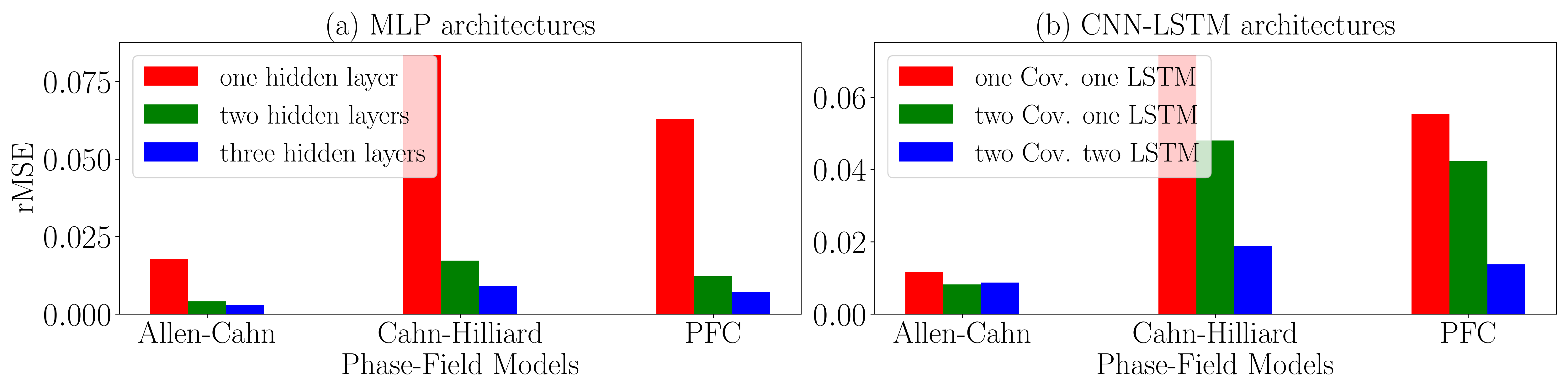}  
  \label{tabl: Allen--Cahn}
\end{subfigure}
\newline
\begin{subfigure}{\textwidth}
  \centering
  \includegraphics[width=\linewidth]{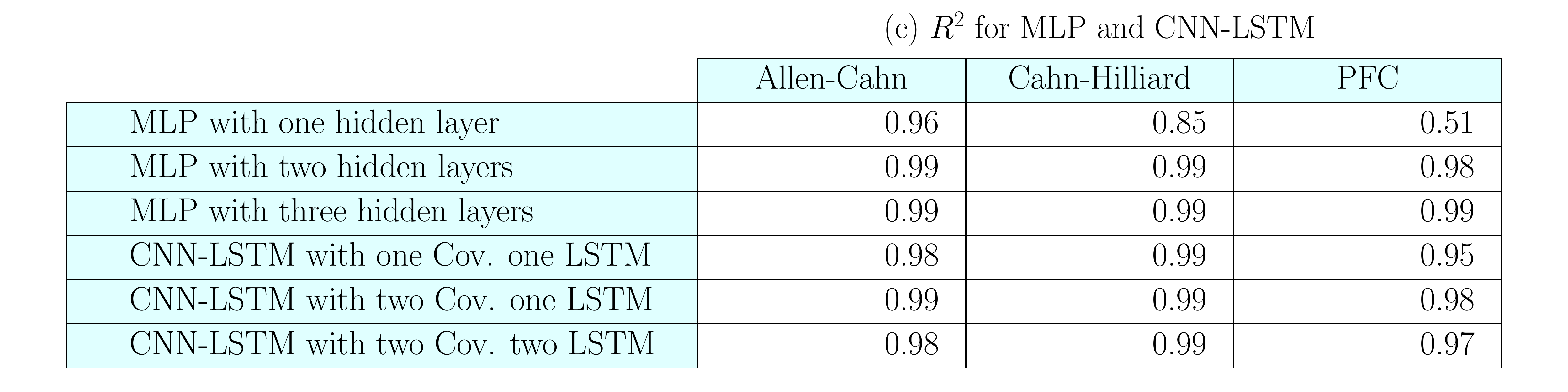}  
  \label{tabl:Cahn--Hilliard}
\end{subfigure}
\newline

\caption{Effect of changing MLP and CNN-LSTM architectures on rMSE and $R^2$. (a) rMSE values obtained by three different MLP architectures, (b) rMSE values obtained by three different CNN-LSTM architectures. (c) $R^{2}$ values for the test set calculated by Equation~\eqref{eq:R2} reported for three different MLP and CNN-LSTM architectures.
}
      \label{fig:bars-compare}
 \end{figure}

Figure~\ref{fig:bars-compare}a) shows the effect of adding layers to our MLP architecture. An MLP network with 64 hidden neurons is expanded to 128/64 and 256/128/64 hidden neurons. It can be seen that adding hidden layers reduces the rMSE and increases the performance of prediction.
Figure~\ref{fig:bars-compare}b) presents the effect of adding CNN and LSTM layers to the CNN-LSTM. Here, we use a single LSTM layer with two configurations for CNN layers: 1) single CNN layer with output filters of size 64, 2) two CNN layers with 128/64 output shape as well as two LSTM layers with 128/64 neurons followed by two CNN layers with 128/64 output sizes. 
Adding convolutional layers increases the performance. However, MLP networks are more sensitive to the choice of architecture than the CNN-LSTM networks. Moreover, the computational cost of training a multi-layer CNN-LSTM is huge compared to a single layer and should be taken into account for large-scale data. It can be roughly concluded that the optimal number of LSTM and CNN layers is $1$ in our CNN-LSTM network. Conversely, the $R^2$ values show less sensitivity to the structural changes in our proposed neural networks, particularly in the CNN-LSTM network (see Figure~\ref{fig:bars-compare}c)).

\begin{figure}
     \centering
     \includegraphics[width=\textwidth]{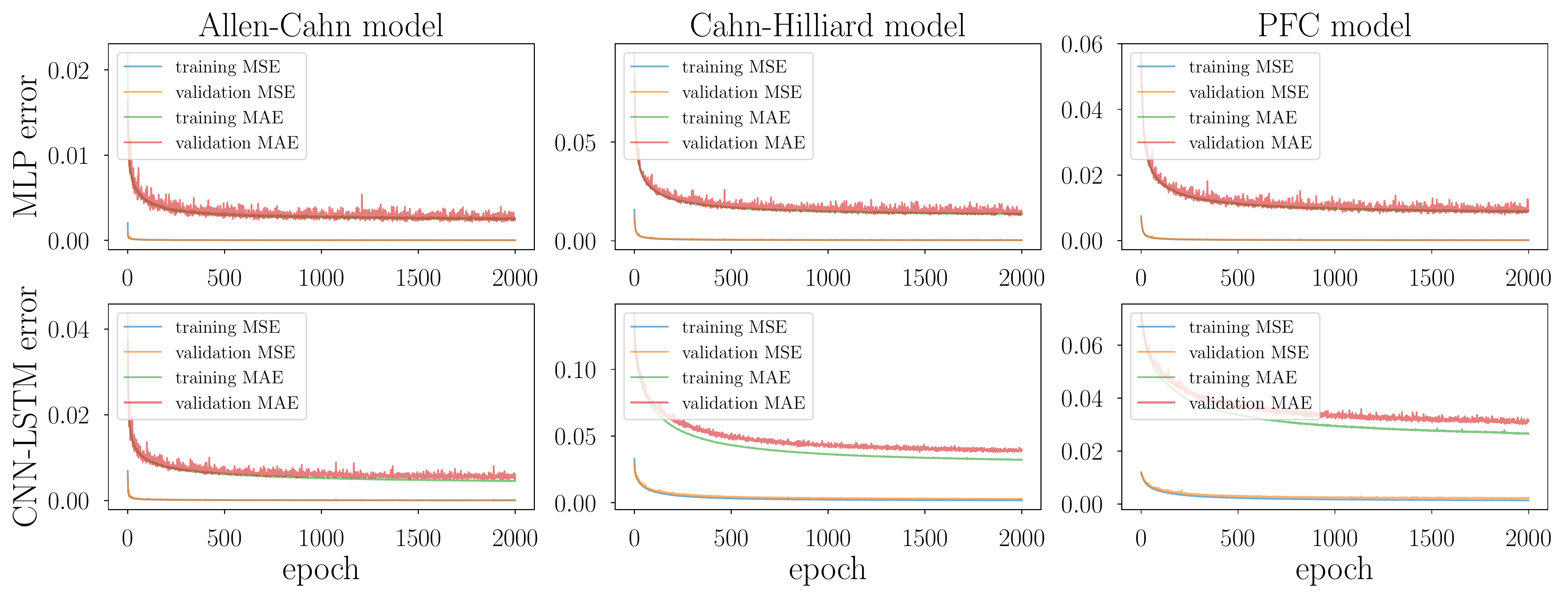}
     \caption{
     Trace of MSE and MAE (see Equations~\eqref{eq:MSE} and  \eqref{eq:MAE}) errors for MLP and CNN-LSTM networks. The blue and green lines represent the errors on
the training sets as a function of epochs, and the orange and red lines correspond
to the errors on the validation sets. Learning curves show that the training and validation curves are very similar for both MSE and MAE errors and they decrease to a point of stability.}
      \label{fig:error_LSTM-CNN-LSTM}
 \end{figure}

To further study the dynamics of the optimization process (training models), the MSE and mean absolute error (MAE) as a function of epochs are given in Figure~\ref{fig:error_LSTM-CNN-LSTM}. 
The MAE is the difference between the original and predicted values. This is calculated by averaging the absolute difference over the data set and is expressed as
\begin{equation}\label{eq:MAE}
    {\mathrm {MAE}}=\frac{1}{n_x\times n_y} \displaystyle \sum_{i=1}^{n_x\times n_y} \lvert U^i_{t} - \widehat{U_t^i} \rvert.
\end{equation}

In order to achieve sufficiently small error, we trained networks for 2,000 epochs with a batch size of 64. However, using approximately $500$ epochs (e.g.~early-stopping \cite{prechelt1998early}) seems adequate for achieving optimal results, particularly for the Allen--Cahn and the Cahn--Hilliard models. Since training CNN-LSTM networks is computationally expensive, using smart early-stopping approaches can help in cases of large data PDE learning tasks.
 
\section{Data-Driven PDEs without Spatial Derivatives Dictionary}\label{CNN}

In this section, we reformulate the problem of learning PDEs as a black-box supervised learning task using convolutional neural network architecture where there is no selection of spatial derivatives and the field $U$ is the only input to our deep learning model. The mathematical representation of data-driven PDE learning task with CNN is
\begin{equation}
\begin{split}
    & U_t(t, x, y)  = F_{\mathrm{CNN}}\left(U(t,x,y)\right) \\
    & F_{\mathrm{CNN}}: \mathbb{R}^{n_x\times n_y} \to  \mathbb{R}^{n_x\times n_y},
\end{split}
\end{equation}
where $n_x$ and $n_y$ are the number of grid points in the $x$- and $y$-directions, respectively. 
We use 
$U$ from our phase-field model simulations to train the CNN.
After successful training of the CNN networks, 
arbitrary initial conditions were chosen for the field $U$ and it was evolved in time by solving $U_t = F_\mathrm{CNN}(U)$ numerically at each grid point.
\begin{figure}
    \centering
    \includegraphics[width=0.99\textwidth]{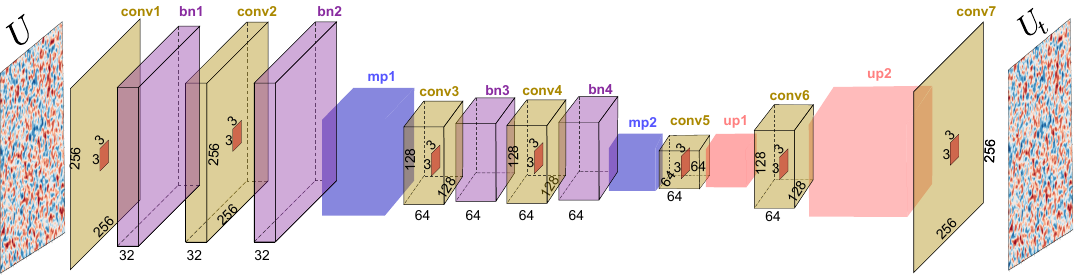}
    \caption{The proposed CNN architecture. The input and output of the CNN are the $U$ and $U_t$ fields, respectively. Input passes through several convolution (\texttt{conv}), batch normalization (\texttt{bn}), max pooling (\texttt{mp}) and up-sampling (\texttt{up}) layers. 
    All the relevant parameters of the network architecture are described in Section~\ref{subsec:cnn_arch}
    }
    \label{fig:cnn_arch}
\end{figure}
\subsection{Convolutional Neural Network (CNN) Architecture}
\label{subsec:cnn_arch}
The CNN network architecture is illustrated in Figure~\ref{fig:cnn_arch}. The full details of the mathematical operations and functionality of each layer are beyond the scope of this paper and can be found in reviews on CNNs such as the one by Rawar and Wang~\cite{rawat2017deep}.
The CNN structure proposed here, similar to the U-net \cite{ronneberger2015u,lapeyre2019training}, resembles the encoding-decoding (auto-encoding) networks. The scalar field which is discretized on $n_x\times n_y$ grid points was fed as the input to the network. In the contracting path, two convolutional layers (\texttt{conv1, conv2} in Figure~\ref{fig:cnn_arch}) with 32 filters each followed by ReLU and batch normalization (\texttt{bn1, bn2}) were applied. 
The kernel size was $3\times 3$ for all the convolutional layers. After the \texttt{bn2} layer, the 2D max pooling operation (\texttt{mp1}) with zero stride (for dimensionality reduction purposes) was applied. The pool size for all the max pooling layers was $2\times2$. The same sub-structure is repeated with 64 filters (\texttt{conv3, bn3, conv4, bn4}) up to the bottleneck unit (output of \texttt{mp2}).
The expansion path consists of two convolutional layers (\texttt{conv5,conv6}) with ReLU units, each followed by an upsampling layer (\texttt{up1, up2}) with the expansion factor of $(2,2)$. 
Finally, at the last convolutional layer (\texttt{conv7}), a linear activation function was used with a filter of size one resulting in an output of shape $n_x \times n_y$. The stochastic gradient decent optimization was applied to find the parameters of the network, where the cost function is the mean absolute error between the network output and $U_t$ from the training set. In total, our CNN network consists of 121,057 trainable parameters.

\subsection{CNN performance for learning PDEs}

\begin{figure}
    \centering
    \includegraphics[width=\textwidth]{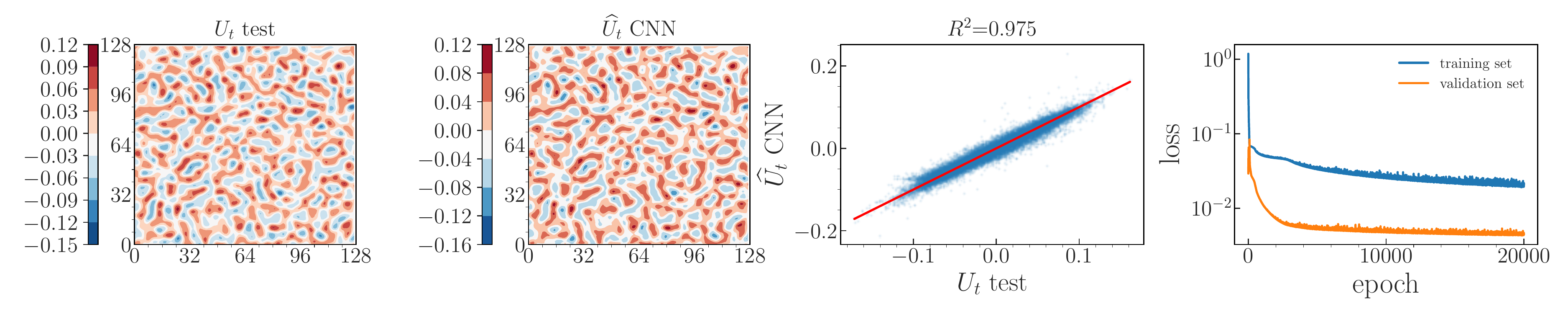}
    \caption{Results using the CNN model trained on the Cahn--Hilliard (Equation~\eqref{eq:modelb}) dataset. The left two panels show the contour of the $U_t$ 
    test set and the corresponding prediction by the CNN. The $\widehat{U_t}$ predictions for all test data as well as the traces of the loss functions are given in the right two panels.
}
    \label{fig:CNN-prediction}
\end{figure}

The phase-field models presented in Section~\ref{Phase-field model}
were used to evaluate the performance of the CNN network. For each model, the total of $n_t$ two-dimensional $U$ and $U_t$ fields were used and randomly split 80:10:10 into training, validation and test sets, respectively. The $U$ and $U_t$ fields from training sets were provided as an input and output to the CNN.
All models were trained for 20,000 epochs and the performance of the network to recover the $U_t$ (learning the RHS of a PDE) on the 
test sets is summarized in Table~\ref{tab:cnn_input_data} in terms of $R^2$ values. The  values indicate that all the trained models performed outstandingly in recovering the PDEs. The contours of $U_t$, and the prediction of the CNN (for the Cahn–Hilliard model, Equation~\eqref{eq:modelb}) are compared in Figure~\ref{fig:CNN-prediction} in the left two panels. The figure shows a qualitative comparison between the original and the data driven models. In addition, all the $U_t$ values in the test set are compared to the CNN predictions. The data lie mostly on the diagonal line indicating good performance. 
\begin{table}
\centering
\begin{tabular}{l|l|l|l}
\hline
2D Model & Allen--Cahn Equation~\eqref{eq:modela} & Cahn--Hilliard Equation~\eqref{eq:modelb} & PFC Equation~\eqref{eq:pfc}   \\ \hline
  $R^2$  &  0.98 & 0.975  & 0.985  \\ 
 \hline
\end{tabular}
\caption{ Our proposed CNN model is  trained for each dataset and the prediction performance on the test (unseen) data are given in the $R^2$ row.
}
\label{tab:cnn_input_data}
\end{table}
The traces of the loss/cost functions during the training phase are also given in the rightmost panel of Figure~\ref{fig:CNN-prediction}. Similar results/plots were obtained for both the Allen--Cahn (Equation~\eqref{eq:modela}) and the PFC (Equation~\eqref{eq:pfc}) model (data not shown).
\subsection{Simulation of Data-Driven PDEs}

In this section, the potential of the proposed method to predict the field $U$ in time and space based on a given initial condition $U_0$ is presented.
For all three phase-field models (Section~\ref{Phase-field model}), we are interested in solving a set of PDEs of the form
\begin{equation}
\begin{split}
    & \frac{\partial U(t,x,y)}{\partial t} = F_\mathrm{CNN} \left(U(t,x,y)\right) \\
    & U(0,x,y) = U_0  \mathrm{; \quad initial \; condition,}
\end{split}
\end{equation}
where the right hand side
is the output (prediction) of the trained CNN networks.

In the following, we used the $U$ fields at $t=2$ (simulation time) as the initial condition ($U_0$) for all the three models. The $U$ field had $n_x \! \times \! n_y = 128 \! \times \! 128$ real values for the Cahn--Hilliard (Equation~\eqref{eq:modelb}) and the PFC (Equation~\eqref{eq:pfc}) models, and $256 \! \times \! 256$ for the Allen--Cahn model (Equation~\eqref{eq:modela}). The different sizes were used to test if there is any size dependence.
At each time ($t$) the $U_t$ values for every grid point were determined from our trained CNN models and $n_x\times n_y$ ODEs were solved using the (stiff) integrator. We used the \texttt{scipy} Adams/BDF method with automatic stiffness detection and switching for time integration~\cite{2020SciPy-NMeth, hindmarsh1983odepack}. Those ODEs were solved up to $t=6$ in our benchmark dataset.

\begin{figure}
    \centering
    \begin{tabular}{c}
        \includegraphics[width=0.78\textwidth]{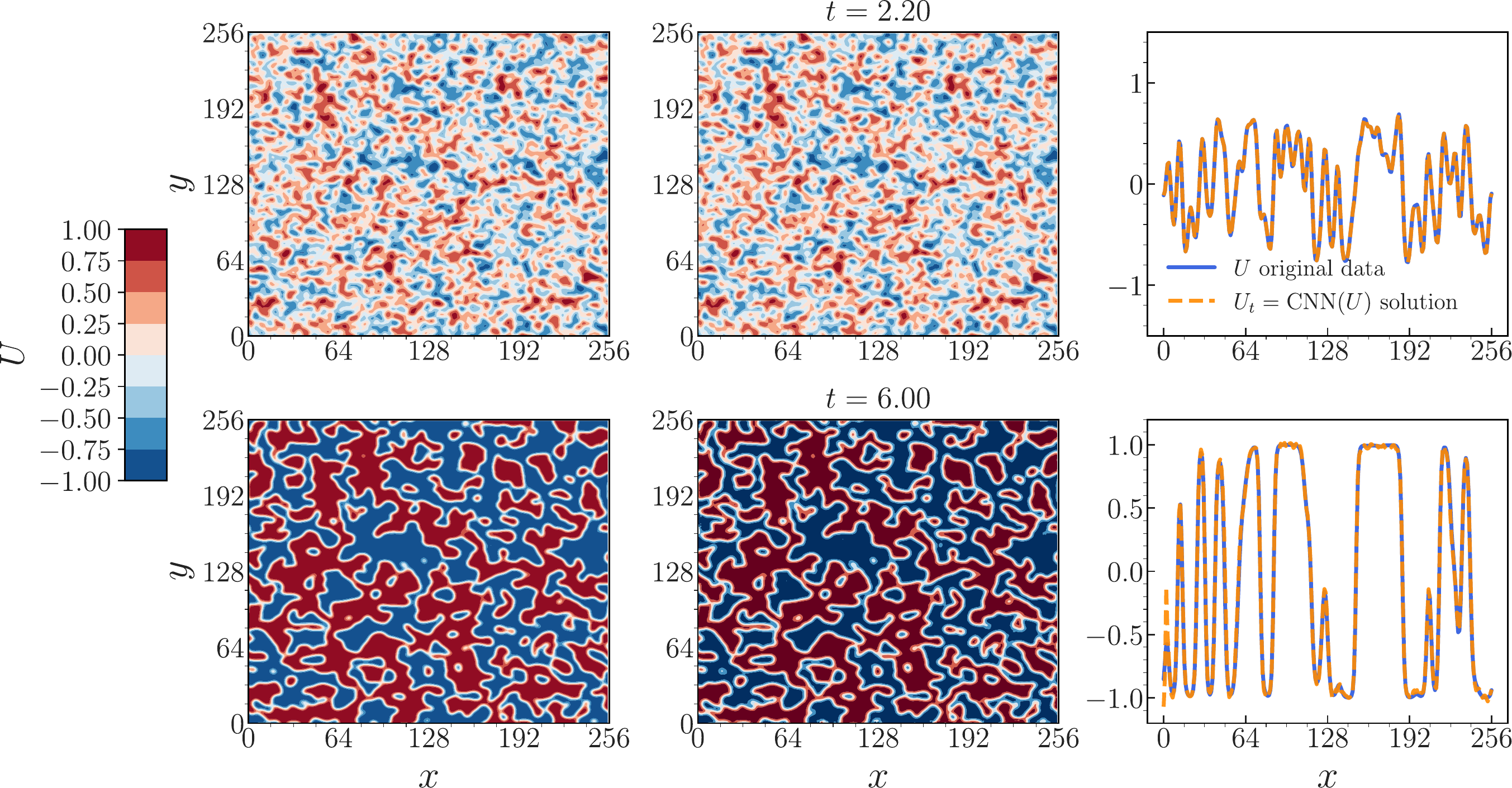} \\
        (a) Allen--Cahn model\\
        \includegraphics[width=0.78\textwidth]{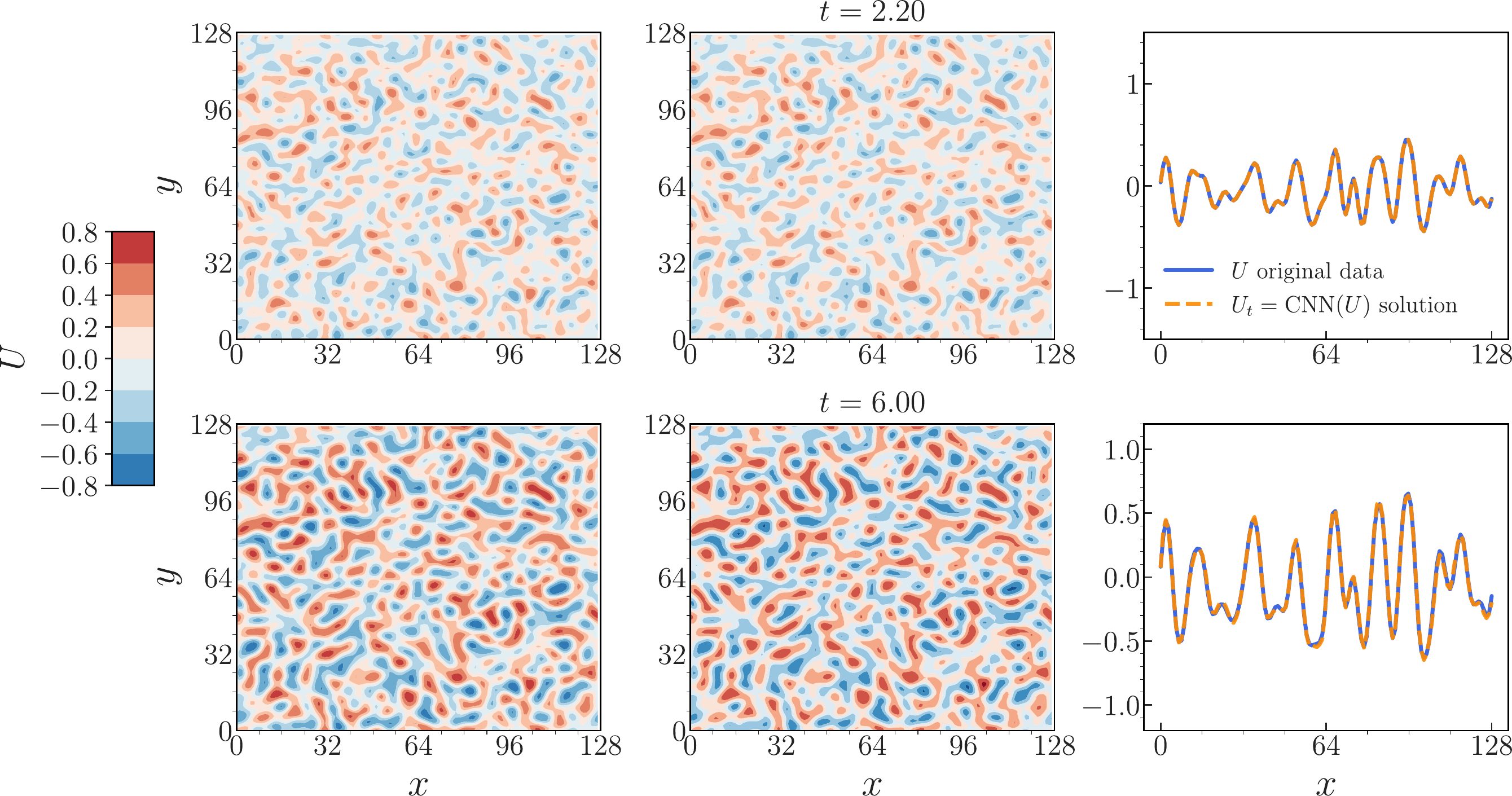} \\
        (b) Cahn--Hilliard model\\
        \includegraphics[width=0.78\textwidth]{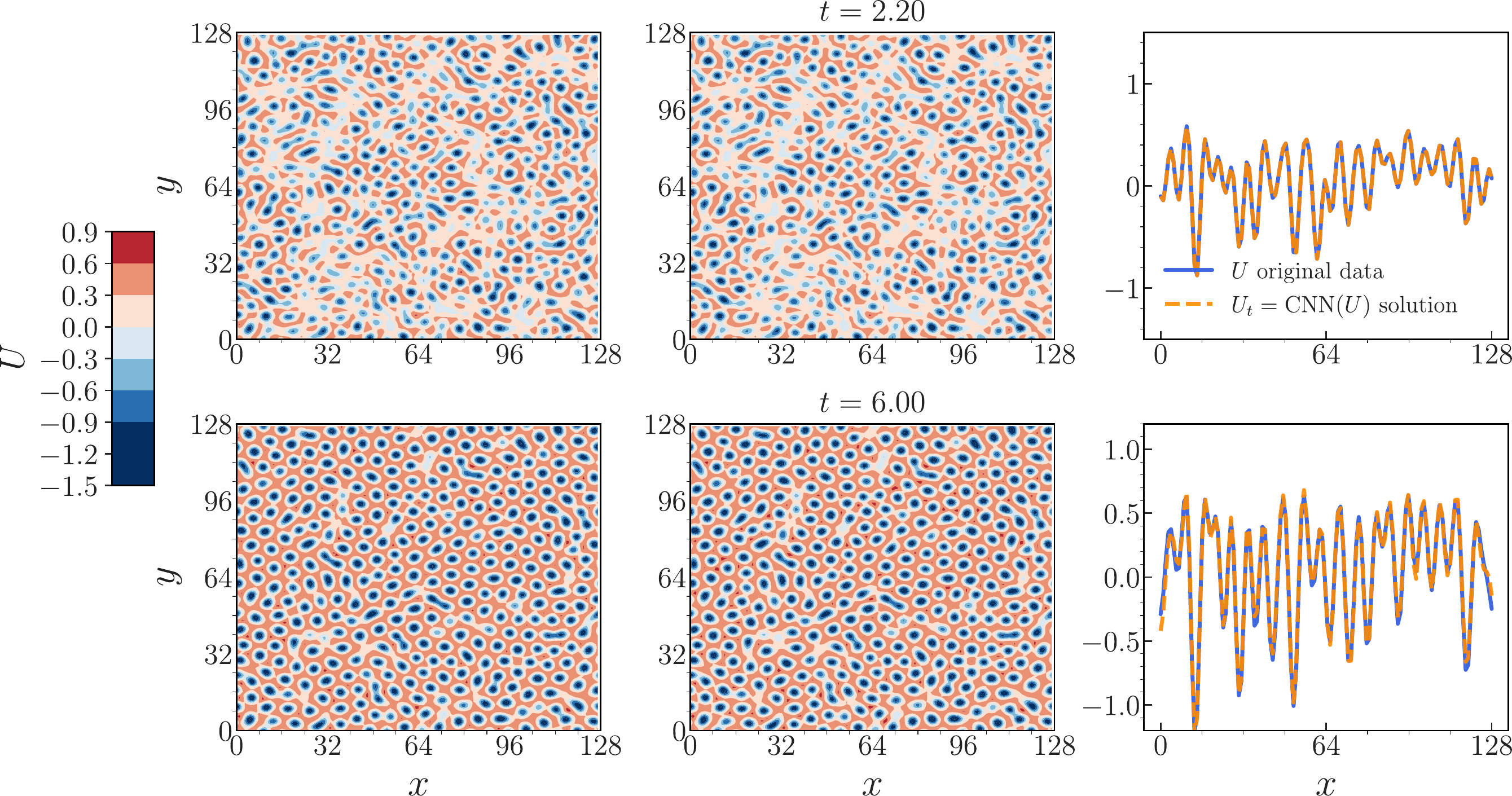} \\
        (c) PFC model
    \end{tabular}
    \caption{Time integration results of the PDEs learned by CNN for (a) the Allen--Cahn (Equation~\eqref{eq:modela}), 
    (b) the Cahn--Hilliard (Equation~\eqref{eq:modelb}) and (c) PFC (Equation~\eqref{eq:pfc}) models at $t=2.2$ and $t=6$. Left panels: $U$ field for original data. Middle panels: $U$ field from simulations of the learned PDEs. Right panel: $U$ values along the centerline $y=n_y/2$ for the original PDEs (solid-lines) and from simulations of the learned PDEs (dashed lines).}
    \label{fig:cnn_time_integ}
\end{figure}

Figure~\ref{fig:cnn_time_integ} shows the solutions of the original and the data-driven PDEs. The contours for $U$ are given for qualitative comparison as well as the $U$ values along the centerline $y\!=\!n_y/2$ for two snapshots at times $t \! = \!2.2$ and $t\!=\!6$.
The results in Figure~\ref{fig:cnn_time_integ} showed that the data-driven PDEs learned by CNN approximate the original dynamics in both quantitative and qualitative manner. 

Finally, we would like to emphasize the following points: 1) The explicit forms of the data-driven PDEs are not known and there is no obvious relation between the functional form of the original and the learned PDEs. Therefore, unlike with the phase-field models, there is no guarantee for existence and uniqueness for the learned PDEs. 2) There are some isolated points in which the $U_t$ predicted values are different from the original models. This discrepancy propagates in time and space, and can lead to finite time blow-up in simulations. This is a known issue in (almost all) machine learning algorithms for time series forecasting where there is no periodicity in time. In the case of no underlying periodicity, it may occur that the 
system trajectories do not span the phase space properly, that is, in such a case the observations are not representative. Such a situation may limit the applicability of the approach to short simulation times. 

\section{Conclusion}

We have presented a data-driven methodology for discovering PDEs from phase-field dynamics.
The well-known Allen--Cahn, Cahn--Hilliard and phase-field crystal models were used as the test cases to predict the underlying equations of motion.

First, we provide an MLP architecture to learn the PDEs where the spatial derivatives are explicitly or implicitly given.
Second, CNN-LSTMs were employed to learn the governing PDEs from coarse-scale local values. 
Third, we proposed a special CNN architecture for cases where there is no information about spatial dependence.
In addition, using numerical integration, we showed how the learned PDEs can be used to predict coarse-scale variables as a function of time and space, starting from given initial conditions. The evolution of the learned and original PDEs showed excellent agreement.
We emphasize that all of the above algorithms yield a black-box-type discovery of PDEs with no obvious connection to the functional form of the physical models.

In general, MLP networks are extremely flexible with data, and PDEs can be learned from various types of data using these networks. More specifically,
they can be used to learn a mapping from a coarse field and its spatial derivatives as the inputs.
However, the performance of an MLP network is greatly affected by the choice of architecture, and also the training of such networks requires the use of spatial derivatives. Along with approximating derivatives, we need to know the derivatives' orders as that is required to train an MLP network.

In CNN networks, however, spatial derivatives are not required, and thus a CNN can be thought of as a finite-difference method capable of estimating derivatives in its first convolution layer.
Moreover, one major advantage in using CNNs is their capability to extract spatial features from inputs. Since LSTMs pass only time information to the layers and keep the missing spatial information from the previous steps, a combination of CNNs and LSTMs can be applied more generally on data with spatial relationships, and, in the current case,  to learn phase-field models. In spite of these advantages, CNN networks are memory intensive and require a large amount of data and several iterations in order to be trained effectively and LSTMs are computationally expensive.
Despite the above limitations, 
we believe that the techniques introduced here offer approaches that are both general and systematic, and provide a basis for future developments.

\section*{Acknowledgments}
Mahdi Kooshkbaghi partially supported by NIH Grant GM133777. Mikko Karttunen thanks the Natural Sciences and Engineering Research Council of Canada (NSERC) and the Canada Research Chairs Program. Computing facilities were provided by SHARCNET (www.sharcnet.ca) and Compute Canada (www.computecanada.ca).


\end{document}